# On profile reconstruction of Euler-Bernoulli beams by means of an energy based genetic algorithm

A. Greco[a], A. Pluchino[b], S. Caddemi[a], I. Caliò[a], F. Cannizzaro[a]

*[a] Department of Civil Engineering and Architecture, University of Catania,*

*Via Santa Sofia 26, 95125, Catania, ITALY.*

*[b] Department of Physics and Astronomy University of Catania and Sezione INFN of Catania*

*Via Santa Sofia 26, 95125, Catania, ITALY*

## Abstract

This paper studies the inverse problem related to the identification of the flexural stiffness of an Euler Bernoulli beam in order to reconstruct its profile starting from available response data. The proposed identification procedure makes use of energy measurements and is based on the application of a closed form solution for the static displacements of multi-stepped beams. This solution allows to easily calculate the energy related to beams modeled with arbitrary multi-step shapes subjected to a transversal roving force, and to compare it with the correspondent data obtained through direct measurements on real beams. The optimal solution which minimizes the difference between measured and calculated data is then sought by means of genetic algorithms.

In the paper several different stepped beams are investigated showing that the proposed procedure allows in many cases to identify the exact beam profile. However it is shown that in some other cases different multi-step profiles may correspond to very similar static responses, and therefore to comparable minima in the optimization problem, thus complicating the profile identification problem.

## 1. Introduction

It is very well known that an important task for an engineer is the detailed knowledge, for every construction typology, of the geometric and mechanical characteristics of the structure and the maintenance of its integrity against natural degradation over time. In the context of health monitoring several studies have been presented in the literature concerning damage detection on beams [1]-[15], frames [16]-[18] and arch structures [19]-[24]. Another important aspect, which has received minor attention in the scientific research, concerns the correct identification of the profiles of existing beams whose visual inspections are not allowed. To this purpose an attempt to identify the stiffness distribution in a structure employing a FEM approach can be found in [25]. The detailed knowledge of the shape, and therefore of the flexural stiffness, of each beam composing the entire structure, is fundamental in order to build a reliable model for the study of either the dynamic or static behavior of the structure under assigned external loads [26].

Within the context of inverse problems focused either on the identification of the presence of damage or of the geometry of the structure, reference can be made to static or dynamic tests.

Dynamic load tests provide global information such as natural frequencies and mode shapes, however, static tests are easily executable and can give useful information without introducing uncertainties due to masses and damping ratios. In the literature there are several studies, dedicated to the identification of both physical and geometrical parameters of the structural systems which perform identification procedures based on dynamic measurements [1][4]-[11][14][20]-[30]. A comprehensive review of damage identification techniques that exploit modal curvature analysis can be found in [31]. Less numerous, although relevant, are the papers proposing static tests [2],[3],[17]-[19].

In this paper an original approach for the identification of the geometrical characteristics of beams, in particular of the shape of their profile, is proposed. Although the proposed procedure can be framed in the context of static identification approaches, usually based on single load distribution tests, it features the same property of inverse identification procedures relying of dynamic tests that is of dealing with global response data. In fact, the proposed objective function to be minimized for the solution of the inverse problem, consists in the total external work related to a concentrated load roving along the beam axis that is directly related to its elastic energy in view of the Clapeyron theorem. Since the total static response of the beam is taken into account in evaluating this work, the considered data provide useful global information for the identification of the beam profile.

The execution of the proposed static test requires therefore the application of a concentrated load of given intensity which must be moved along a beam axis grid measuring, at each position, the transversal displacement below the load. The total external work can therefore be evaluated as the sum of the work done by the force for the transversal displacements in the current, and in all the positions previously assumed by the load.

The procedure proposes to compare these static data with those provided by a closed form solution for the evaluation of transversal displacements according to a multi-stepped beam model, proposed in [32],[33], in order to seek, within this beam typology, the profile which better suits the measured data.

The availability of an analytical explicit equation for the evaluation of static displacements in multi-stepped beams allows to explore the responses related to all the possible profiles rapidly and to find the best one by means of an optimization procedure based on the use of genetic algorithms.

In the applicative section of the paper many examples of the capability of the proposed procedure of identifying the correct profile of beams are provided. Furthermore, the robustness of the procedure is studied and, according to the precision of the input data, some counter-intuitive static responses in multi-stepped beams are highlighted together with the related effects leading to a non uniqueness of the solution of the inverse profile identification problem.

## 2. A flexural stiffness model with multiple singularities

In this section a model of Euler-Bernoulli beam showing multiple singularities is presented, and its capability of describing discontinuities in the response functions is shown. The Euler-Bernoulli beam model subjected to external transversal loads $\overline{q}(x)$, and accounting for a spatial variable flexural stiffness $E(x)I(x)$, is governed by the following differential equations with respect to the spatial abscissa $x$ spanning from 0 to the beam length $L$ :

$$\left[E(x)I(x)v''(x)\right]'' = \overline{q}(x) \tag{1}$$

In Eq. (1) the apex indicates differentiation, with respect to the abscissa *x,* and $v(x)$ is the transverse deflection function. The flexural stiffness is considered variable according to the following law:

$$E(x)I(x) = E_o I_o \left[1 - \sum_{j=1}^{m} \left(\beta_j - \beta_{j-1}\right) U\left(x - x_j\right)\right] \tag{2}$$

which accounts for $m$ abrupt decreases in the flexural stiffness with respect to a maximum reference value $E_o I_o$. The parameters $\beta_j$, $j = 1,\ldots,m$, represent the flexural jumps and $x_j$ are the relevant singularity positions, respectively. The definition of the flexural stiffness requires the adoption of the well-known generalized Dirac's delta functions $\delta\left(x - x_j\right)$ and Heaviside $U\left(x - x_i\right)$.

The external transversal load distribution $\overline{q}(x)$ can be generalized since can model both continuous $\overline{q}_0(x)$ and discontinuous loads, included transversal point loads $\overline{P}_r$ at $x_{P_r}$, $r = 1,\ldots,n_\blacklozenge$, as follows:

$$\overline{q}(x) = \overline{q}_0(x) + \sum_{r=1}^{n_\blacklozenge} \overline{P}_r \, \delta(x - x_{P_r}) \tag{3}$$

For simplicity, by considering the dimensionless coordinate $\xi = x/L$, and indicating with the prime the differentiation with respect to $\xi$, the governing differential equation of the Euler-Bernoulli beam given by Eq. (1), by accounting for the singularities introduced in Eqs. (2),(3), takes the following form:

$$\left\{\left[1 - \sum_{j=1}^{m} \left(\beta_j - \beta_{j-1}\right) U\left(x - x_j\right)\right] u''(\xi)\right\}'' = q(\xi) \tag{4}$$

Eq. (4), expressed in term of the normalized function $u(\xi) = \dfrac{v(\xi)}{L}$ , accounts for generic external loads $\overline{q}(x)$, given by Eq. (3) referred to the normalised abscissa $\xi$, by means of the normalised transversal load parameter $q(\xi)=\dfrac{\overline{q}(\xi)}{E_0 I_0}L^3$. In Eq. (4) the property $\delta\left[L(\xi-\xi_i)\right]=\left(1/L\right)\delta(\xi-\xi_i)$ of the Dirac's delta distribution has been exploited.

Integration of Eq.(4) leads to the following general form of the deflection function:

$$u(\xi)=c_1 f_1(\xi)+c_2 f_2(\xi)+c_3 f_3(\xi)+c_4 f_4(\xi)+f_5(\xi) \tag{5}$$

where the constants $c_i, i=1,...,4$, represent the integration constants to be obtained as a function of the boundary conditions, while the $f_i\left(\xi\right),\ i=1,...,5$ functions are given by:

$$\begin{aligned}
&f_1(\xi)=1 \qquad ; \qquad f_2(\xi)=\xi\\
&f_3(\xi)=\xi^2+\sum_{j=1}^{m}\beta_j^{\,*}(\xi-\xi_j)^2\,U(\xi-\xi_j)\\
&f_4(\xi)=\xi^3+\sum_{j=1}^{m}\beta_j^{\,*}(\xi^3-3\xi_j^2\xi+2\xi_j^3)U(\xi-\xi_j)\\
&f_5(\xi)=q^{[4]}(\xi)+\sum_{j=1}^{m}\beta_j^{\,*}\left[q^{[4]}(\xi)-q^{[4]}(\xi_j)\right]U(\xi-\xi_j)+\\
&\qquad -\sum_{j=1}^{m}\beta_j^{\,*}q^{[3]}(\xi_j)(\xi-\xi_j)U(\xi-\xi_j)
\end{aligned} \tag{6}$$

In Eqs. (6) the $^{[p]}$ indicates the $p$-th integral of the function and the following parameters have been introduced:

$$\beta_j^{\,*}=\frac{\beta_j}{1-\beta_j}-\frac{\beta_{j-1}}{1-\beta_{j-1}} \tag{7}$$

When the beam is subject to a constant distributed vertical load $q_0$, the $f_5(\xi)$ function in Eq.(6) becomes:

$$\begin{aligned}
f_5(\xi)=&\frac{q_0\xi^4}{24}+\sum_{j=1}^{m}\beta_j^{\,*}\frac{q_0}{24}\left(\xi^4-\xi_j^4\right)U(\xi-\xi_j)+\\
&-\sum_{j=1}^{m}\beta_j^{\,*}\frac{q_0\xi_j^3}{6}(\xi-\xi_j)U(\xi-\xi_j)
\end{aligned} \tag{8}$$

while in the case of a single concentrated load the $f_5(\xi)$ function in Eqs.(6) becomes:

$$f_5(\xi) = P\frac{(\xi-\xi_P)^3}{6}U(\xi-\xi_P) + \sum_{j=1}^{m}\beta_j^*\left[\frac{(\xi-\xi_P)^3}{6}U(\xi-\xi_P) - \frac{(\xi_j-\xi_P)^3}{6}U(\xi_j-\xi_P)\right]PU(\xi-\xi_j) + \\ -\sum_{j=1}^{m}\beta_j^* P\frac{(\xi_j-\xi_P)^2}{2}(\xi-\xi_j)U(\xi_j-\xi_P)U(\xi-\xi_j) \tag{9}$$

where the normalized concentrated load is given by $P = \dfrac{\bar{P}L^2}{E_0 I_0}$

In the following Table 1, the four integration constants for some significant boundary conditions are reported.

Table 1. Boundary conditions and corresponding integration constants

| Boundary conditions | Mathematical conditions | Integration constants |
| --- | --- | --- |
| Simply supported beam | $u(0)=0;\quad u''(0)=0;$ <br> $u(1)=0;\quad u''(1)=0$ | $c_1=0;\quad c_2=\dfrac{f_5''(1)}{f_4''(1)}f_4(1)-f_5(1);$ <br> $c_3=0\quad c_4=-\dfrac{f_5''(1)}{f_4''(1)}$ |
| Clamped clamped beam | $u(0)=0;\quad u'(0)=0;$ <br> $u(1)=0;\quad u'(1)=0$ | $c_1=0;\quad c_2=0;$ <br> $c_3=\dfrac{\dfrac{f_5'(1)}{f_4'(1)}f_4(1)-f_5(1)}{f_3(1)-\dfrac{f_3'(1)}{f_4'(1)}f_4(1)};$ <br> $c_4=-\dfrac{\dfrac{f_5'(1)}{f_4'(1)}f_4(1)-f_5(1)}{f_3(1)-\dfrac{f_3'(1)}{f_4'(1)}f_4(1)}\dfrac{f_3'(1)}{f_4'(1)}-\dfrac{f_5'(1)}{f_4'(1)}$ |
| Cantilever beam | $u(0)=0;\quad u'(0)=0;$ <br> $u''(1)=0;\quad u'''(1)=0$ | $c_1=0;\quad c_2=0;$ <br> $c_3=\dfrac{f_5'''(1)f_4''(1)}{f_3''(1)f_4'''(1)}-\dfrac{f_5''(1)}{f_3''(1)};\quad c_4=-\dfrac{f_5'''(1)}{f_4'''(1)}$ |

It is worth pointing out that this closed form solution of the Euler-Bernoulli beam under static loads requires the enforcing of four boundary conditions only irrespectively of the number of the along beam steps in the flexural stiffness. This solution will be exploited in the context of an inverse problem aiming at identifying the best step distribution of the flexural stiffness which approximate the static response of a real beam. To this purpose, an energy cumulative parameter able to collect the static response of the beam subjected to a roving static concentrated load will be employed, as better shown in the next section.

## 3. Beam discretization and objective function for profile identification

In order to identify the profile of a generic beam with a variable flexural stiffness, it is convenient to consider a simplified model with $m$ abrupt equally spaced changes in the stiffness that correspond to an arbitrary number $N_P$ of uniform segments (of course $m \leq N_P$) of the beam. This kind of beam is called "multi-step". In Figure 1 an example of a multi-step beam discretization is reported, with $N_P = 10$ and $m = 6$. The concentrated load $P$ roving along the beam is represented as an arrow which progressively changes its location, applied in the centre $\xi_P$ of each segment, where a device for the measurement of transversal displacement is also present.

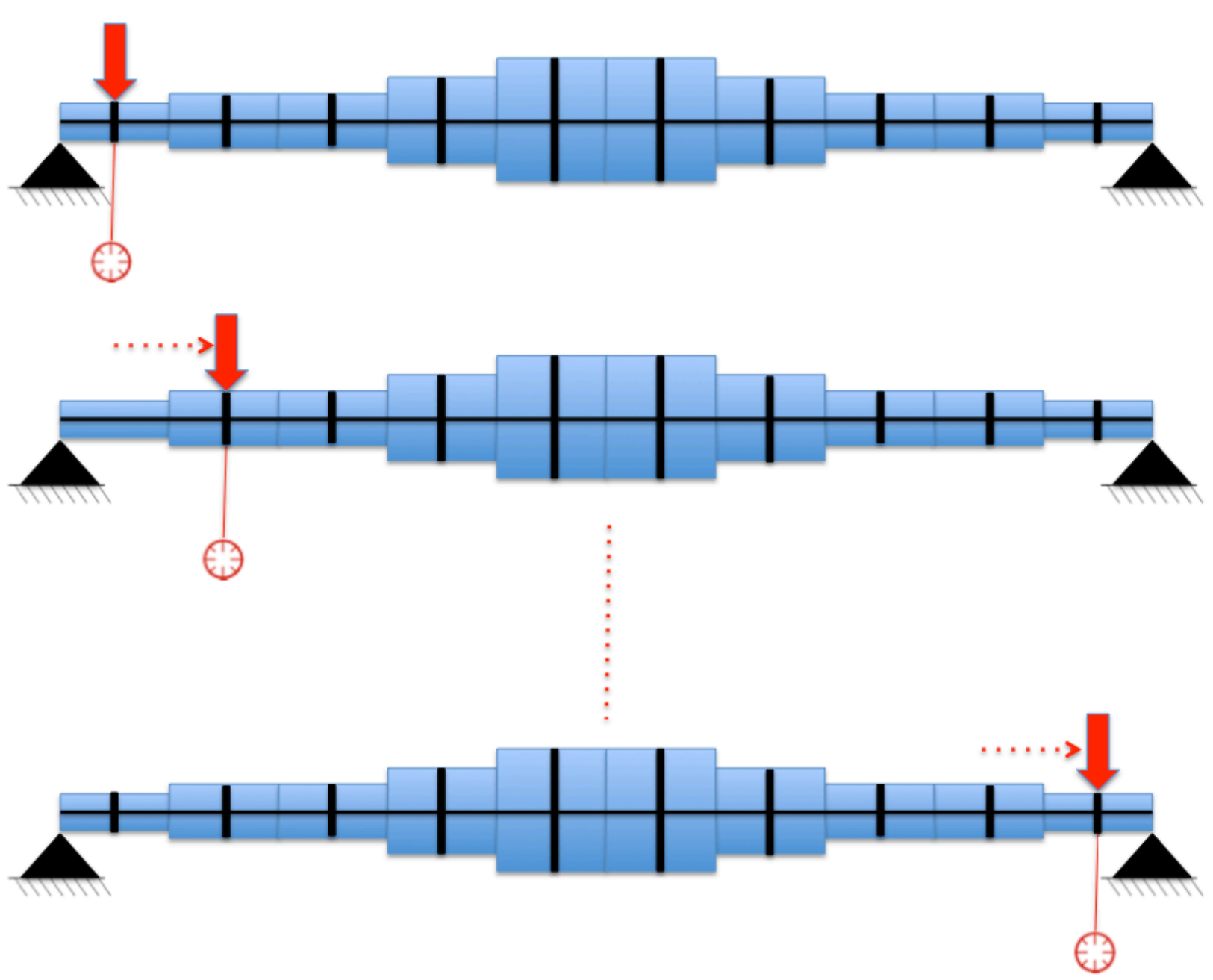

Figure 1. Discretized stepped beam with roving concentrated load

The closed form expression, provided in the previous paragraph, for the transversal displacement of a stepped beam subjected to a concentrated load, allows a straightforward evaluation of the total external work $W$ done by a concentrated load roving along the beam axis. In fact, $W$ can be expressed as the summation of the products of the concentrated load times the transversal displacement of the cross section crossed by the load itself. Once a suitable sequence of $N_p$ points with interval $\Delta\xi$ is chosen and the position $\xi_P$ of the roving load is considered ranging from $\xi = 0$ to $\xi = 1$ along the grid, the total external work as function of an increasing number $n_P \leq N_P$ of visited positions $\xi_{Pk}, k = 1,\ldots,n_P$, can be expressed, in view of Eqs. (5), (6) and (9) as follows:

$$W(n_P) = P\sum_{k=1}^{n_P}\left\{\begin{array}{l} c_1 + c_2\xi_{P_k} + c_3\left[\xi_{P_k}^2 + \sum_{j=1}^{m}\beta_j^*(\xi_{P_k} - \xi_j)^2 U(\xi_{P_k} - \xi_j)\right] + \\ +c_4\left[\xi_{P_k}^3 + \sum_{j=1}^{m}\beta_j^*(\xi_{P_k}^3 - 3\xi_j^2\xi_{P_k} + 2\xi_j^3)U(\xi_{P_k} - \xi_j)\right] \end{array}\right\} \tag{10}$$

According to the expression in Eq.(10), the direct contribution of the external work of the roving load to the transversal displacement at the same cross section where the load is applied disappears and its influence is related only to the integration constants.

Following Eq.(10), the total external work $W(n_P)$ can be calculated for any $n_P \leq N_P$ by adding the products of the concentrated load times the measured transversal displacement of the cross section crossed by the load itself. Depending on the beam profile, the values of both $N_P$ and $m$, and the chosen boundary conditions, a given trend of $W(n_P)$ as function of the roving load position $n_P$ can be obtained.

Figure 2 shows some illustrative examples of the trend of the total external work for multi-step beams with $N_P = 10$ and the three considered boundary conditions. As it can be observed the curves are always increasing, as expected; those related to simply supported, Figure 2(a), and clamped-clamped beams, Figure 2(c), present an inflection point since the vertical displacements towards the right end of the beams assume decreasing values. On the contrary the total external work for multi-step cantilever beams, Figure 2(b), has increasing slope.

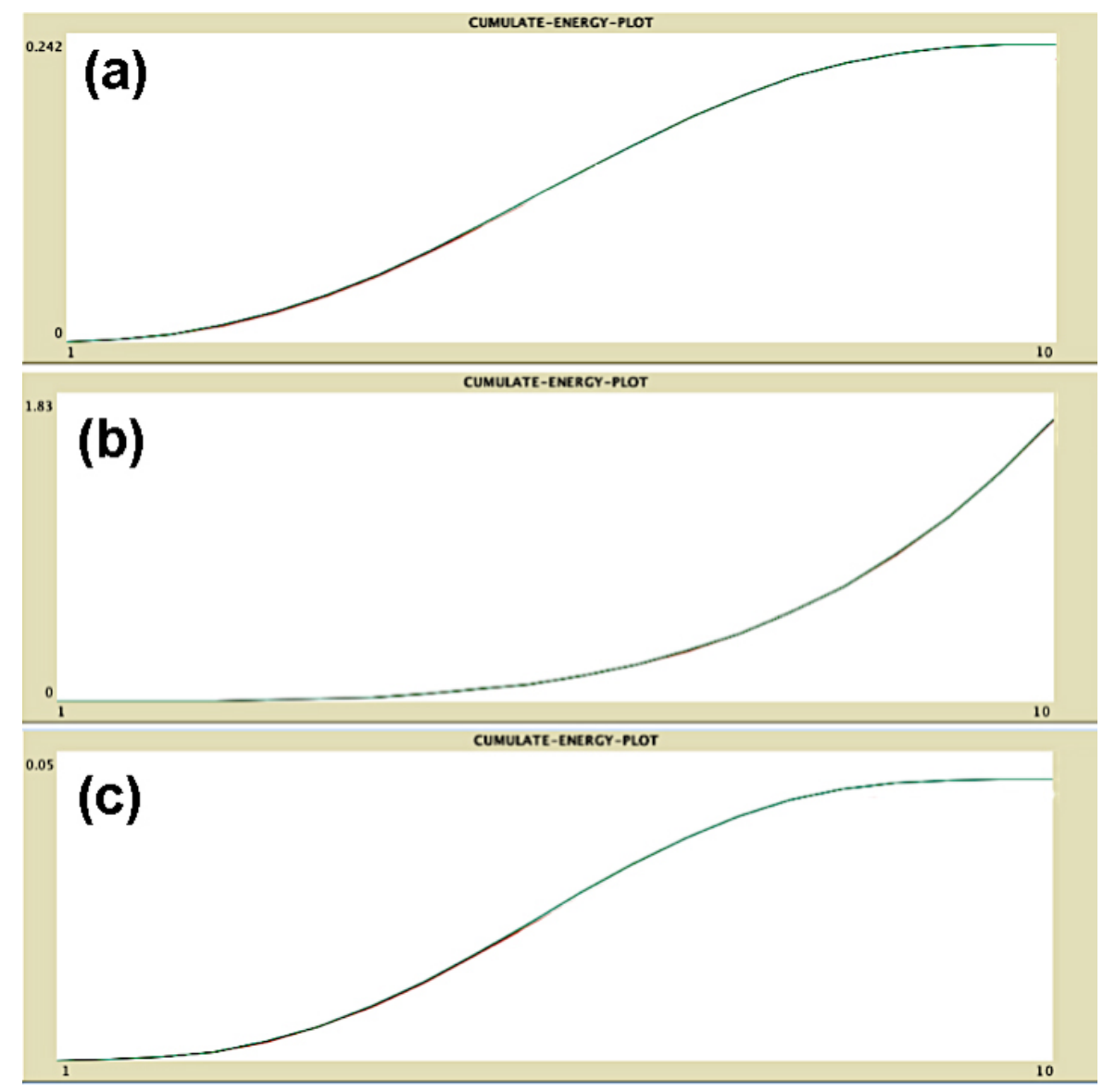

Figure 2. Total external work trend for (a) simply supported beam, (b) cantilever beam, (c) clamped-clamped beam

At this point, taking into account all these prescriptions, it is possible to summarize the procedure for the inverse problem regarding the identification of the profile of a generic real beam starting from its total external work $W_r$, which is made of the following fundamental steps:

1. Define an arbitrary number $N_P$ of uniform segments with the same length along the real beam;
2. Apply a roving load at the mid-point of each segment defined at point 1 and measure the transversal displacement below the load;
3. Calculate the real total external work $W_r(n_P)$ (function of the roving load position $n_P$);
4. Generate several different multi-stepped profiles of the beam, with $m \leq N_P$ (like that one in Fig.1);
5. For each new profile assigned at point 4, calculate the total external work $W_c(n_P)$ as function of the roving load position $n_P$;
6. Evaluate the following objective function $O_b$ which depends, for each position of the roving load, on the differences between the value of the real total external work (point 3) and that of the corresponding calculated one (point 5), normalized to the total external work $W_0(n_P)$ of the beam with constant reference stiffness $E_o I_o$:

$$O_b = \sum_{n_p=1}^{N_p} \left[ \frac{W_r(n_p) - W_c(n_p)}{W_0(n_p)} \right]^2 \tag{11}$$

7. Find the multi-stepped profile that minimizes the objective function exploring different multi-step profiles by means of an optimization procedure based on genetic algorithms, as described in the next paragraph. If the inverse problem results well-conditioned, the found profile should coincide with that of a given real multi-stepped beam or should represent the best multi-step approximation of a given real smooth profile.

## 4. Optimization procedure and the Genetic Algorithm based Software

A "genetic algorithm" is an adaptive stochastic method that mimic the Darwinian evolution, based on an opportune combination of random mutations and natural selection, in order to numerically find optimal values of some specific function. The algorithm acts over a population of $P$ chromosomes, corresponding to potential solutions of the considered problem, by iteratively applying the "survival of the fittest" principle. In such a way it produces a sequence of new generations of chromosomes that evolves towards a stationary population where the large majority of surviving solutions do coincide and approach as much as possible the real solution of a practical problem [34].

Concerning the problem of reconstructing the profile of a generic beam with a variable flexural stiffness, discretized in $N_p$ uniform segments, a given chromosome is coded as a string of integer numbers, where each number (called gene) is related to the stiffness of the corresponding segment, directly related to the beam cross section. Therefore, a generic chromosome $C_i$ of the population ($i = 1, \ldots, P$) can be coded in the following string:

$$C_i \quad (c_1, c_2, c_3, \ldots c_k, \ldots, c_{N_p})$$

with $c_k \in [0, c_{\max}]$, where $c_{max}$ indicates the maximum decrease with respect to the reference stiffness $E_o I_o$ ($c_k=0$ means that the corresponding segment has the reference stiffness, $c_k=c_{max}$ is associated to zero flexural stiffness). In particular, recalling the definition of the singularity parameter $\beta$ introduced in Eq.(4), the stiffness of the $k$-th segment of the discretized beam will be $E_o I_o (1 - \beta_k)$, with $\beta_k = c_k / c_{max}$.

The overall number of possible different chromosomes is thus $P_{max} = (c_{max} + 1)$ . The task of the genetic algorithm is to explore the space of all the possible chromosomes, in search of the multi stepped profile of the beam which maximizes an opportune "fitness function", here defined as:

$$f(C_i) = F_{max} - O_b(C_i) - \gamma G(C_i) \qquad (12)$$

where $O_b(C_i)$ is the objective function (11) calculated for a generic chromosome $C_i$ and $G(C_i)$ is a cost function, defined as:

$$G(C_i) = \frac{1}{N_P - 1} \sum_{i=1}^{N_P - 1} \left| \beta_{i+1} - \beta_i \right| \qquad (13)$$

whose purpose is to penalize chromosomes with a great average stiffness variation between contiguous segments of the discretized beam and to favor the chromosomes characterized by a smoother profile. Thus, the addition of the cost function favors chromosomes corresponding to beams with a more realistic profile. The parameter $\gamma$ modulates the relative weight of this cost with respect to the objective function. Finally, $F_{max}$ is an arbitrary constant, chosen great enough to have $f(C_i) > 0$ for every possible value of $O_b(C_i)$ and of the cost function $G(C_i)$. In the following it will be set $F_{max} = 10$ without loss of generality (it should be noticed that, presenting the numerical results, $G(C_i)$ – which typically always remains strictly greater than zero – will not be considered in the fitness visualization, in order to have $f(C_i) = F_{max}$ when the correct chromosome is recovered).

The fitness value associated to each chromosome represents its probability of survival, under the pressure of the natural selection process. In the following it is explained more in detail how this process does work.

Starting from the initial population of $P$ chromosomes (with, typically, $P = 100$), randomly chosen among the $P_{max}$, a new generation is created from the old one, where chromosomes that have a higher fitness score are more likely to be chosen as "parent" than those that have low fitness scores. The selection method adopted in this paper is called "tournament selection", with a tournament size of three: this means that groups of 3 chromosomes are drawn randomly from the old generation, and the one with the highest fitness in each group is chosen to become a parent. Either one or two parents are chosen to create children: with one parent, the child is simply a clone of the parent; with two parents, the process is the digital analogue of sexual recombination – the two children inherit part of their genetic material from one parent and part from the other (crossing-over). Once the new generation is created, there is also a chance that random mutations will occur at level of the single genes $c_k$ of the child chromosomes, and some of them will be changed into new ones (always chosen in the interval [0, $c_{\max}$]).

By iteratively repeating this process several times, chromosomes with the highest fitness will be progressively selected in the space of all the possible combinations and will quickly spread among the population reducing the diversity of the individuals, until (almost) only one of them will survive: hopefully, that one with the maximum fitness. Of course, it is frequent for the dynamics to remain trapped into local maxima of fitness therefore it is convenient to launch the genetic algorithm many times (events), each time starting from a different initial population, in order to gain more chances to reach the global maximum of fitness.

## 5. Numerical applications

In this section several numerical applications of the procedure previously described, are presented. All the applications have been developed using an original software code, which runs in NetLogo [35], a freeware multiplatform environment with an owner high level programming language and with a very ductile and versatile user interface.

NetLogo platform was natively developed for agent-based simulations and for modelling complex systems behaviour. The idea is to harness the power of the NetLogo graphical user interface and the versatility of its agent-oriented programming language in order to create an user-friendly original software for the automatic profile reconstruction of beams.

In the next subsection, the investigation of the inverse problem concerning profile identification of multi-stepped target beams with different boundary conditions is presented. In all the addressed examples, the target chromosome, i.e. the one corresponding to the target beam, has always been sought to be the one with the best fitness (equal to $F_{max}$) after some runs of the genetic algorithm.

Successively, a deeper numerical analysis of the inverse problem, devoted to estimate the robustness of the procedure, will be performed. First, it will be shown that the introduction of an increasing instrumental error in the calculation of the transversal displacements below the roving load (which propagates, in turn, on the calculation of the total external work), progressively reduces the performance of the genetic algorithm. Then, it will be also shown that beams with multi-step profiles, sensibly different from the target one, may exhibit very similar static responses (i.e. may correspond to chromosomes with quite high fitness) while beams with multi-step profiles, very similar to the target one, may exhibit very different static responses (i.e. may correspond to chromosomes with lower fitness). This latter numerical evidence makes the problem very difficult to be governed, particularly in presence of several steps. However, this is a result important to highlight for the development of profile identification procedures.

*5.1 Identification of multi-stepped beam profiles*

In this paragraph several beams with both different multi-step profiles and boundary conditions are analysed in order to test the performance of the identification procedure when the target chromosome does coincide with the target beam. For all of them, a discretization with $N_P$ = 10 uniform segments with the same length is considered; the identification procedure aims therefore at identifying the flexural stiffness of these uniform portions of the beam. A unitary vertical load is applied alternatively at the middle cross sections of each segment starting from the left end of the beam. For each segment, ten different values for the flexural stiffness have been considered, starting from the reference stiffness $E_oI_o$ and decreasing it in abrupt changes. The application of the genetic algorithm deals therefore with chromosomes of 10 genes whose integer values can vary from $c_k = 0$ (reference stiffness) to $c_{max} = 9$ and the space of the total number of possible chromosomes has therefore a size $P_{max} = 10^{10}$.

It is important to notice that in all these tests the target chromosome, corresponding to the desired multi-stepped beam, can be directly inserted as an input string in the NetLogo interface, then the software automatically calculates the transversal displacements below the roving load and the consequent total external work, which represents the reference data to be matched in the inverse problem. After several tests the optimal set of parameters for the genetic algorithm have been chosen as following: population of $P$=100 chromosomes evolving for 100 generations, cross-over and mutation rates respectively set to 80% and 1%, cost function weight $\gamma = 0.06$ (the low value of $\gamma$ intends to favour, here, the minimization of the objective function with respect to the cost one). As already anticipated, the maximum fitness parameter $F_{max}$ is set to 10. Either symmetric or arbitrary distributions of the uniform segments have been considered for beams with the three boundary conditions (simply supported, clamped-clamped, cantilever) described in paragraph 2. Table 1 reports some examples of the considered beams and the correspondent target-chromosome.

In Figure 3 the software user interface at the beginning of a typical simulation is shown. In particular, the simply supported beam *SS1,* shown in Table 2, has been considered. The multi-step profile of the beam (where each segment is just represented with a vertical bar placed in the corresponding centroid) is reported in the central upper part of the interface, together with the plots of the total external work for both the target beam (in black) and the one with constant reference stiffness (in red). In the left part of the interface, buttons and sliders for setting the input parameters of the beam (included the target chromosome) are visible, while in the right part those concerning the input parameters of the genetic algorithm are present.

Table 2 Tested multi-stepped beam profiles

| Multi-step target beam | Target-chromosome |
|---|---|
| SIMPLY SUPPORTED *SS1* | [9 7 5 3 0 0 3 5 7 9] |
| SIMPLY SUPPORTED *SS2* | [2 3 4 3 4 5 6 3 2 1] |
| CLAMPED-CLAMPED *CL1* | [1 3 5 7 9 9 7 5 3 1] |
| CLAMPED-CLAMPED *CL2* | [2 3 4 3 4 5 6 3 2 1] |
| CANTILEVER *CA1* | [0 1 2 3 4 5 6 7 8 9] |
| CANTILEVER *CA2* | [2 3 4 3 4 5 6 3 2 1] |

Finally, at the bottom of the central-right part of the interface, the output window and the plots of both the average fitness and the diversity of chromosomes as function of the generations are reported.

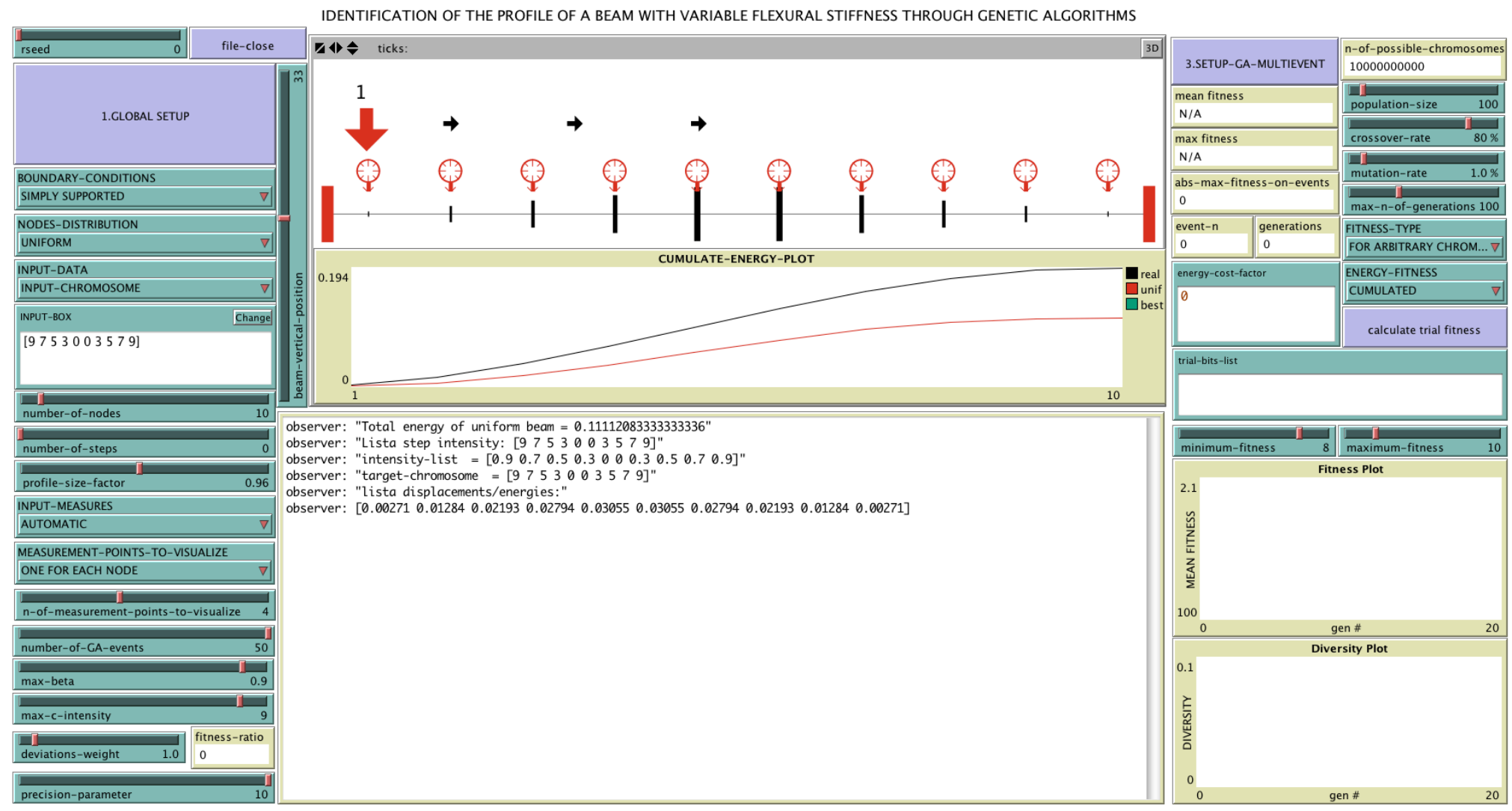

Figure 3. Software user interface at the beginning of a typical simulation.

The diversity evaluates the average 'disagreement' among all the chromosomes in the population of the current generation and it is based on the well-known Hamming distance, which essentially counts the number of genes, which have different values in any couple of chromosomes.

Running the algorithm when the number of generations increases, the fitness converges towards its (local or global) maximum value while the diversity goes to zero, meaning that the winning chromosome tends to spread among the population. In Figure 4, a typical behaviour of both the fitness and the diversity plots during a single event simulation is reported.

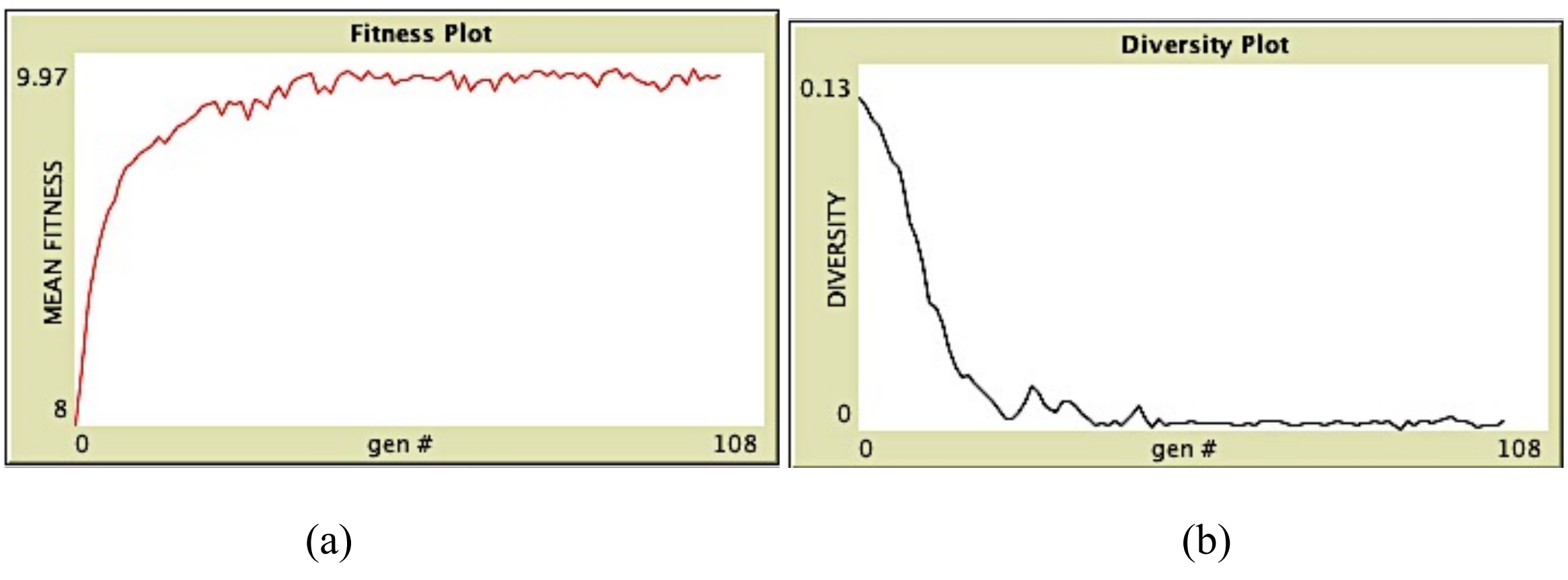

(a) (b)

Figure 4. Behavior of fitness (a) and diversity (b) during a single event simulation.

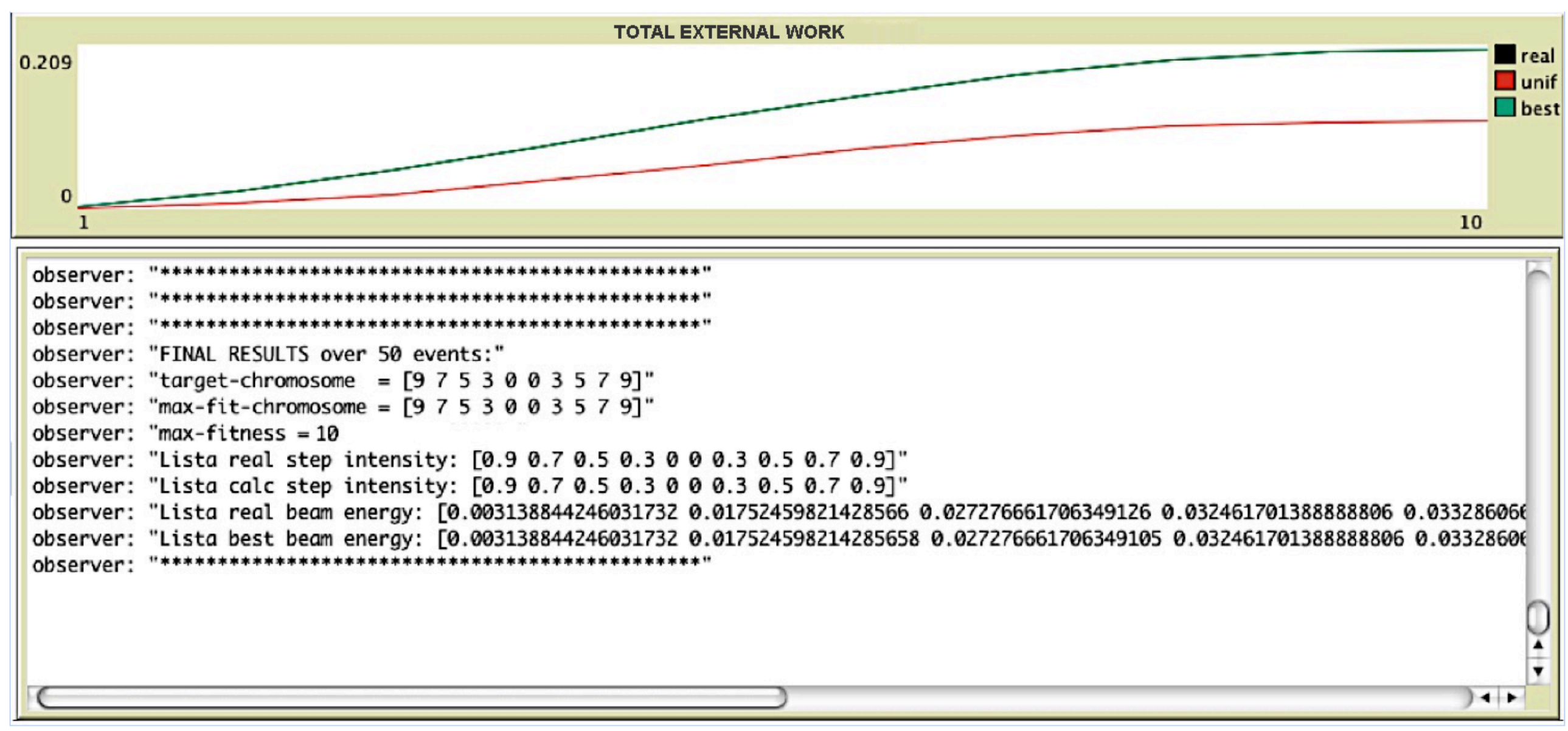

Figure 5. Software user interface at the end of the simulation after 50 events for the simply supported symmetric beam of Table 2.

As visible in Figure 5, at the end of a complete simulation with 50 events (each starting from a different random population of $P$ chromosomes) the target chromosome $C_T \equiv [9\ 7\ 5\ 3\ 0\ 0\ 3\ 5\ 7\ 9]$ has been correctly recovered, in the space of all the possible ones, with a maximum fitness value for the winning chromosome $C_W$ equal to $f\,(C_W) = 10$, i.e. with no errors. This is confirmed by the total external work plot, where a perfect correspondence between the target beam curve and the one related to the best chromosome (in green) can be appreciated.

Analogous behaviours, with a correct retrieval of the target profile, have been obtained for all the other cases of multi-stepped beams shown in Table 2.

Consider, just to show another example, the multi-stepped cantilever beam *CA1* shown in Table 2. For example, Figure 6 refers to the multi-stepped cantilever beam *CA1* and shows the final outcome of a 50 events simulation, together with the corresponding total external work plot: again, the target chromosome $C_T \equiv [0\ 1\ 2\ 3\ 4\ 5\ 6\ 7\ 8\ 9]$ has been correctly recovered with a fitness value $f\,(C_W) = 10$.

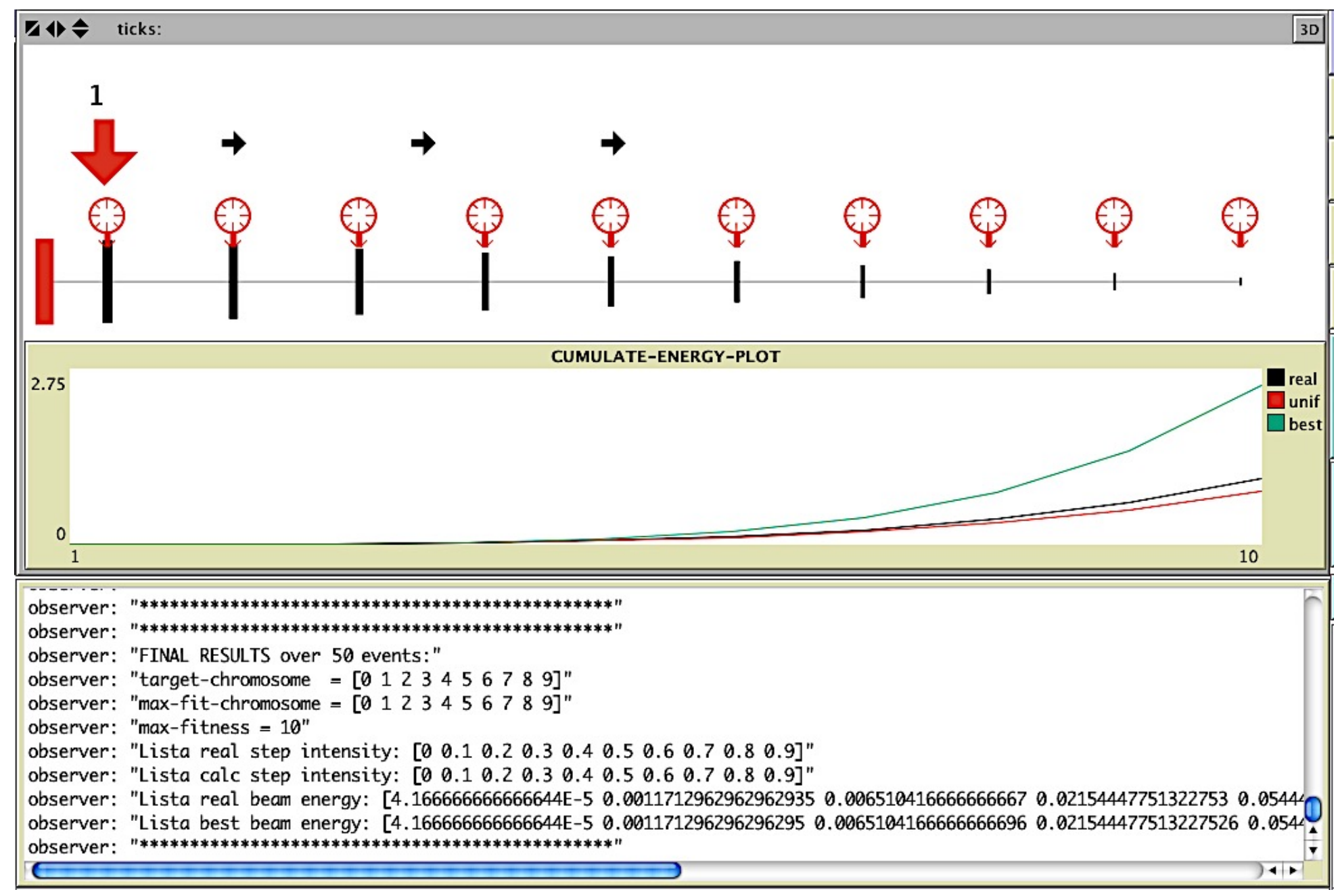

Figure 6. User interface for the cantilever beam with progressively decreasing stiffness.

*5.2 Robustness of the procedure*

As it has been illustrated in the previous paragraph, the proposed procedure is able to provide the exact solution when the beam, whose profile must be reconstructed, has a multi-step shape. This happens because, in this case, the winning chromosome can return the correct total external work profile with a very good numerical approximation, therefore allowing the correct identification of the target multi-step beam. It is now interesting to study how the results presented in the previous subsection depend on the precision of the input and also how robust is the multi-step identification procedure with respect to variations of some genes in the target chromosome.

In order to address these issues, consider again the multi-stepped cantilever beam with progressively decreasing flexural stiffness, whose target chromosome is $C_T \equiv [0\ 1\ 2\ 3\ 4\ 5\ 6\ 7\ 8\ 9]$. The latter was correctly recovered by a winning chromosome $C_W$ with fitness value $f\,(C_W)= F_{max}$. The total external works $W_c(C_T)$ and $W_c(C_W)$ for the chromosomes $C_T$ and $C_W$, calculated below the roving load, are reported in Table 3. It is worth to remind that for each uniform segment the total external work contain the sum of the external works done by the force for the transversal displacements in the current, and in all the positions previously assumed by the load.

Table 3 Total external work for chromosomes $C_T$ and $C_W$

| Load position | $W_c(C_T)$ | $W_c(C_W)$ |
| --- | --- | --- |
| 1 | 4.166666666666644E-5 | 4.166666666666671E-5 |
| 2 | 0.0011712962962962935 | 0.001171296296296294 |
| 3 | 0.006510416666666667 | 0.006510416666666652 |
| 4 | 0.02154447751322753 | 0.021544477513227498 |
| 5 | 0.054441633597883614 | 0.05444163359788358 |
| 6 | 0.11647172619047616 | 0.11647172619047616 |
| 7 | 0.22258217592592588 | 0.22258217592592575 |
| 8 | 0.3922412367724868 | 0.3922412367724866 |
| 9 | 0.6507991071428569 | 0.6507991071428567 |
| 10 | 1.0321129298941796 | 1.0321129298941791 |

If these total external work profiles are compared with a precision of 16 floating point digits, some very small discrepancies in the single values of the two energy arrays, due to unavoidable numerical approximation in the calculus, can be appreciated.

These discrepancies bring to a fitness value $f(C_W)$= 9.9999999999999999 which, of course, can be well approximated with $f(C_T)$=10.

In order to investigate on the effects of these discrepancies on the solution of the inverse problem considering as input data the total external work, a progressive error in the values is introduced, for example by eliminating a given increasing number of final digits in the entries of the input energy array.

In the following the values of the fitness, with the relative winning chromosome, for several values of decreasing precision in the energy entries are reported in Table 4.

Table 4 Fitness and winning chromosome for different number of floating point digits in the energy entries

| n° floating point digits | Fitness $f(C_W)$ | Fitness error (%) | Winning chromosome $C_W$ |
|---|---|---|---|
| 12 | 9.9999999996341023 | $10^{-10}$ | [0 1 2 3 4 5 6 7 8 9] |
| 10 | 9.9999995604706754 | $10^{-7}$ | [0 1 2 3 4 5 6 7 8 9] |
| 8 | 9.9999944898779582 | $10^{-6}$ | [0 1 2 3 4 5 6 7 8 9] |
| 5 | 9.9988653213175636 | $10^{-3}$ | [0 1 2 3 4 5 6 7 8 9] |
| 4 | 9.9747776590120348 | $10^{-2}$ | [0 1 1 5 1 1 8 4 6 6] |

The observation of Table 4 shows that, for errors lower than $10^{-4}$, the retrieval of the correct multi-step profile is ensured, with a corresponding fitness approximately equal to $f(C_W) = F_{max}$. On the other hand, when the measurement errors go beyond that threshold, the winning chromosome identifies a multi-step profile which is completely different from the correct one, with a decrease in the corresponding fitness $f(C_W)$ which is lower than before but, anyway, not so much as expected by looking at $C_W$'s genes.

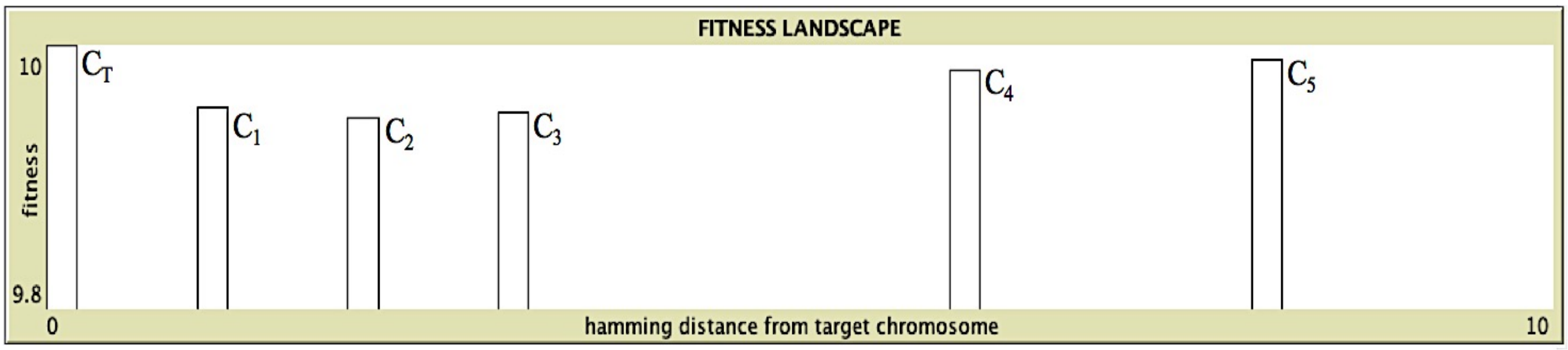

Figure 7. Fitness landscape of chromosomes as function of their Hamming distance from the target chromosome $C_T$.

This finding stimulated a further investigation about the robustness of the multi-stepped beam identification with respect to variations of some genes in the target chromosome.

In this respect, to have a robust identification process, the decrease in the $f\,(C_W)$ fitness value with respect to $F_{max}$=10 should be proportional to the diversity between $C_W$ and $C_T$. Unfortunately, as it will be immediately shown, this is not the case.

As already clarified, the calculation of the diversity among chromosomes is based on the concept of Hamming distance $H(C_i,C_j)$. The latter parameter measures the number of genes which have different values in any couple of chromosomes $C_i$ and $C_j$. Of course, if $C_i = C_j$, $H = 0$, while if the chromosomes are completely different $H = N_P$ (here $N_P$ =10). In Figure 7, a fitness landscape diagram where the fitness of any arbitrary chromosome $C_k$ can be plotted as function of its Hamming distance $H(C_k,C_T)$ from the target chromosome $C_T$, is shown. On the extreme left side one finds $C_T$, at distance $H(C_T,C_T)$=0; then, for increasing distances, other chromosomes are reported, each one with its fitness, in Table 5.

Table 5 Fitness and Hamming distance from the target chromosome for arbitrary chromosomes

| Considered chromosome $C_i$ | Hamming distance $H(C_i,C_T)$ | Fitness $f\,(C_i)$ |
|---|---|---|
| $C_T \equiv$ [0 1 2 3 4 5 6 7 8 9] | $H(C_T,C_T) = 0$ | $f\,(C_T)$= 10 |
| $C_1 \equiv$ [0 1 2 4 4 5 6 7 8 9] | $H(C_1,C_T) = 1$ | $f\,(C_1)$= 9.9531 |
| $C_2 \equiv$ [0 1 2 4 4 5 6 8 8 9] | $H(C_2,C_T) = 2$ | $f\,(C_2)$= 9.9448 |
| $C_3 \equiv$ [0 0 2 4 4 5 6 8 8 9] | $H(C_3,C_T) = 3$ | $f\,(C_3)$= 9.9492 |
| $C_4 \equiv$ [0 0 4 0 4 4 7 7 8 7] | $H(C_4,C_T) = 6$ | $f\,(C_4)$= 9.9806 |
| $C_5 \equiv$ [0 1 1 5 2 0 8 1 0 3] | $H(C_5,C_T) = 8$ | $f\,(C_5)$= 9.9892 |

It can be noticed that, against the intuition, changing the value of a small number of genes with respect to the target chromosome, as done for examples with chromosomes $C_1$, $C_2$ and $C_3$ (corresponding to beams with only a slight change in the value of the stiffness for a small number of uniform segments), provides a significant reduction of the fitness ($f < 9.96$), related to a consistent variation in the respective total external work.

Table 6 Total external work vectors corresponding to assigned chromosomes

| $W_c(C_1)$ | $W_c(C_T)$ | $W_c(C_5)$ |
| --- | --- | --- |
| 4.166666666666686E-5 | 4.166666666666644E-5 | 4.166666666666671E-5 |
| 0.001171296296296294 | 0.0011712962962962935 | 0.0011712962962962942 |
| 0.006510416666666657 | 0.006510416666666667 | 0.006504622962962963 |
| 0.02155439814814812 | 0.02154447751322753 | 0.021412037037037052 |
| 0.05470949074074071 | 0.054441633597883614 | 0.05434375000000018 |
| 0.11771180555555544 | 0.11647172619047616 | 0.11695254629629659 |
| 0.22598495370370353 | 0.22258217592592588 | 0.22333912037037074 |
| 0.3994733796296294 | 0.3922412367724868 | 0.3934525462962965 |
| 0.6640034722222219 | 0.6507991071428569 | 0.6531817129629636 |
| 1.0539085648148143 | 1.0321129298941796 | 1.0307065145502652 |

This is confirmed, for example, by the comparison between the total external work vectors corresponding to $C_T$ and $C_1$ shown in Table 6 where, as expected, the differences become more pronounced towards the right free end of the beam.

On the other hand, considering chromosomes such as $C_4$ and $C_5$, with a higher Hamming distance from $C_T$ (i.e. corresponding to significantly different multi-stepped beam profiles), the related fitness results to be closer to the maximum value $F_{max}$ ($f$ >9.98), as also confirmed by the comparison between the total external work vectors corresponding to, for example, $C_T$ and $C_5$ (see again Table 6). This means that this kind of solutions act as local maxima which trap the genetic algorithm dynamics, making very difficult to reach the global maximum and therefore to achieve the correct solution with $f = 10$.

The previous considerations demonstrate that it is typically not possible to retrieve the correct target chromosome, and therefore to identify the correct real beam from its total external work profile, when its fitness does not coincide, at a given level of numerical approximation, with the maximum value $F_{max}$. This is for example the case of real beams with smooth profile. In fact, for those beams, the solution of the inverse problem would contemplate the retrieval of the multi-stepped beam which better approximates the real one. But the unavoidable, even small, discrepancies between the total external work profile of the real beam and that one of its multi-stepped version, imply that it does not exist a global maximum in the fitness landscape. Therefore, this makes it impossible to identify the correct target chromosome, being the landscape full of local maxima of fitness, corresponding to chromosomes completely different from the target one, which attract the dynamics of the genetic algorithm leading it away from the desired solution.

It should be noticed that this scenario is independent from the adopted discretization $N_P$ of the real beam profile, therefore the identification problem cannot be circumvented by increasing the level of approximation of the real beam through multi-stepped profiles with a greater number of uniform segments.

It is also worth noticing that the described considerations, regarding the non-uniqueness of the solution of the inverse problem, also apply if the number of input data is increased (for example evaluating the transversal displacements in all the nodes when the applied force moves along the beam). This circumstance lies on the peculiar behaviour of the considered problem according to which different multi-step profiles correspond to very similar static responses.

## 6. Conclusions

The present paper investigates on the possibility of identifying the profile (and therefore the flexural stiffness distribution) of real beams making use of measured static data. In particular, in order to take into account a measure able to characterize the global static response of the beam, reference is made to the total external work produced by a concentrated force moving along the beam axis. The total external work measured on real beams, whose profile must be reconstructed, is then compared to those obtained for arbitrary multi-stepped beams by means of a closed form solution. An original application of a tailored genetic algorithm allows selecting the multistep profiles which better suits the measured data. Many applications of the proposed procedure are presented together with a detailed investigation on the robustness of the procedure aimed at the individuation of the convenience and drawbacks of its applicability. In particular, it has been demonstrated that even slight changes in the value of the stiffness for a small number of uniform segments in a beam cause significant differences in the total external work. Therefore it is not always possible to identify a unique correct target chromosome when its fitness does not coincide, at a given level of numerical approximation, with the maximum possible value. However, the identification of the beam's profile could perhaps be performed by means of more selective hybrid strategies that, taking into account different global measures, could allow finding the correct solution. In this context, the study here presented represents a first important step toward the solution of the complex problem of the beam's profile identification.